# Discrete Chaotic Sequence based on Logistic Map in Digital Communications

V. H. Mankar[1], T. S. Das[2] and S. K. Sarkar[3]

[1]Dept. of Electronics & Telecommunication, Govt. Polytechnic, Nagpur

[2,3]Dept. of Electronics & Telecommunication, Jadavpur University, Kolkata

[1]vijaymankar@yahoo.com, [2]tirthasankardas@yahoo.com, [3]sksarkar@etce.jdvu.ac.in

*Abstract-The chaotic systems have been found applications in diverse fields such as pseudo random number generator, coding, cryptography, spread spectrum (SS) communications etc. The inherent capability of generating a large space of PN sequences due to sensitive dependence on initial conditions has been the main reason for exploiting chaos in spread spectrum communication systems. This behaviour suggests that it is straightforward to generate a variety of initial condition induced PN sequences with nice statistical properties by quantising the output of an iterated chaotic map. In the present paper the study has been carried out for the feasibility and usefulness of chaotic sequence in SS based applications like communication and watermarking.*

## I. INTRODUCTION

In the past years, there has been a wider interest in the study of non-linear dynamical systems among the researchers worldwide. The chaotic systems have found applications in diverse fields such as population dynamics, meteorology, hydrodynamics, chemical engineering, cryptography, coding, encryption, pseudo random number generator, spread spectrum digital communication, etc [1-3]. The random like behaviour of chaotic system made it attractive in spite of being deterministic. A chaotic system generates a set of aperiodic signals with a "noise-like" and broad power spectrum. Moreover, system is very much sensitive to its initial conditions. A slight difference in initial conditions will produce totally different sequences.

In a direct sequence DS-SS communication system, sets of non-correlated binary sequences with good auto-correlation and crosscorrelation properties, such as *m*-sequence, Gold sequences and Kasami sequences are used as spreading codes. Maximum length sequences are generated by means of an *m*-stage shift register with feedback with maximum period equal to $2^m$-1. Although *m*-sequences has excellent delta like autocorrelation properties, it has poor cross-correlation properties. Selecting a pair of m-sequences, called preferred m-sequences and summing them modulo-2 generate Gold sequences. It has good cross-correlation as compare to *m*-sequences. Although Gold sequences comes as an alternative to *m*-sequences but it suffers from the drawback like limitation of fixed period length etc [2].

## II. PRELIMINARIES

A chaotic system is a deterministic nonlinear dynamical

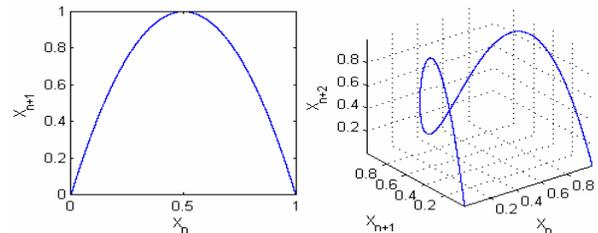

Fig. 1. 2D Phase space (Xn Vs Xn+1) and 3D Phase space (Xn Vs Xn+1 Vs Xn+2)

system whose states change with iterations in a deterministic way.

The logistic map is a one-dimensional discrete-time non-linear system exhibiting quadratic non-linearity. The logistic map is given by the function $f : [0,1] \rightarrow \Re$ defined by

$$f(x) = \mu\, x\, (1\text{-}x) \qquad (1)$$

which is expressed in state equation form as

$$x_{n+1} = f(x_n) = \mu\, x_n\, (1\text{-}x_n) \qquad n = 0,1,2,.., \qquad (2)$$

where $x_n \in (0, 1)$ and $\mu \in (0, 4)$. $\mu$ is known as the *control parameter or bifurcation parameter* .Here $x_n$ is the state of the system at time *n*. $x_{n+1}$ denotes the next state and *n* denotes the discrete time. Repeated iteration of *f* gives rise to a sequence of points $\{x_n\}_{0}$, known as an *orbit*.

Let us briefly goes through some of the interesting dynamic properties exhibited by the logistic map.

We observe the following properties of logistic maps:

- *f* has maximum of $\mu$/4 at $x = 0.5$
- For any *x* between 0 and 0.5 will have symmetric counterpart between 0.5 and 1. Hence $f(x)_{max}$ lies at x = 0.5
- *f* maps [0, 1] back into [0, 1] for $0 \le \mu \le 4$. If *x* ever exceeds unity, the iterations diverge to $-\infty$.

The behaviour of the logistic map is sensitive to the value of $\mu$. The various characteristic features exhibited by varying different values of $\mu$ are discussed below:

### A. *Case I:* $(0 \le \mu < 1)$

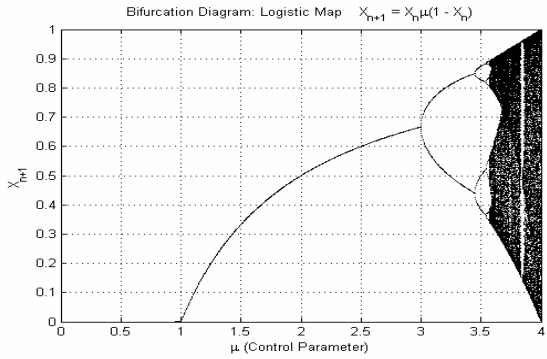

Fig. 2. Bifurcation Diagram

The system always converges to $x = 0$. Thus, all $x$ in the interval $0 \leq \mu < 1$ *attract* to solution $x^* = 0$. They lie in the basin of attraction. This can very well understood through graphical representation of $x_{n+1}$ Vs $\mu$ known as bifurcation diagram as shown in fig. 2.

### B. Case II: $(1 \leq \mu \leq 3)$

From $\mu > 1$ it ceases to converge to 0 and converges to different stable point given by the solution $x^* = 1 - 1/\mu$ (also known as point attractor/ period-1 cycle) till $\mu \leq 3$. The solution is said to become *repeller*. Refer fig. 2 for different values of $\mu$ in this range.

### C. Case III: $(3 < \mu \leq 3.56)$

The attractor at $1 - 1/\mu$ ceases to converge to a fixed point and becomes unstable i.e. it oscillates between two limits. This is shown in fig. 3(a) where for $\mu = 3.2$ it oscillates approximately between 0.513 and 0.7995. We observe the bifurcation of stable point (fig. 2) from $\mu > 3$ and is called *period -2 cycle*. Here $x_{n+2} = x_n$. This happens when $df/dx < -1$. As we increase $\mu$ further at about $\mu = 3.4495$ another period doubling (*period – 4 cycle*) occurs. This happens when $df^2/dx < -1$ where $f^2(x) = f(f(x))$. Refer the fig. 3(b) for $\mu = 3.5$ where

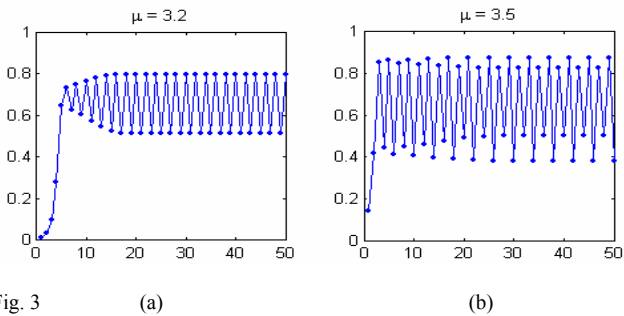

Fig. 3    (a)                    (b)

it oscillates approximately between four limits i.e. 0.3828, 0.8269, 0.5009 and 0.8750.

This different period doubling is shown in fig. 4 which gives rise to another phenomena known as universality. Here $\mu_1 = 3$, $\mu_2 = 3.4495$, $\mu_3 = 3.5441$, $\mu_4 = 3.5644$, $\mu_5 = 3.5688$, $\mu_6 = 3.5697 \ldots \mu_\infty = 3.5699 \ldots$

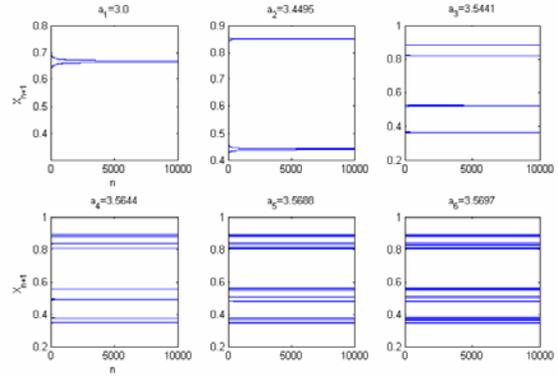

Fig. 4 Period Doubling

Thus, with $\mu$ slightly bigger than 3.54, the iterations oscillate between 8 values, then 16, 32, etc. The lengths of the parameter intervals which yield the same number of oscillations decrease rapidly; the ratio between the lengths of two successive such bifurcation intervals approaches the Feigenbaum constant which is an example of a period-doubling cascade. Feigenbaum discovered that

$$\delta_n = \lim_{n \to \infty} \frac{\mu_n - \mu_{n-1}}{\mu_{n+1} - \mu_n} = 4.669 \ldots \quad (3)$$

This is known as Feigenbaum constant which tells that distance between successive bifurcations in the period doubling shrinks by this constant value.

### D. Case IV: $(3.5669 < \mu \leq 4)$

Most values of $\mu$ in this range produce chaos, which has infinite period. There are also an uncountable number of initial points $x_0$ which give totally aperiodic trajectories; no matter how long the time series generated by $f(x)$ is iterated, the pattern never repeats. The sequences generated are extremely sensitive to the initial condition (seed). This is "chaos" which has been defined as chaos is characterized by long-term aperiodic behaviour in a deterministic system, which exhibits sensitive dependence to initial conditions. Fig. 5 illustrates this phenomenon. Thus, large sources of almost uncorrelated signals are possible as with slight change in the initial condition will produce completely different signal. Moreover, in most cases, the initial condition cannot be deduced from a finite length of the sequence.

Now the question is how to measure sensitive dependence to initial condition? Lyapunov exponent that is discussed as below gives the answer:

Let the two different sequences are iterated with initial condition $x_0$ and $x_0 + \delta_0$ where $\delta_0$ is very small.

Then $x_n = f^n(x_0)$,         $x_n + \delta_n = f^n(x_0 + \delta_0)$,

The sequence $| \delta_n |$ is therefore

$| \delta_n | = | f^n(x_0 + \delta_0) - x_n |$

= $| f^n(x_0 + \delta_0) - f^n(x_0) |$ which describes the "distance" between two orbits. Note that,

$| \delta_n | \approx | \delta_0 | \, e^{\lambda n}$ for small $\delta_n$.

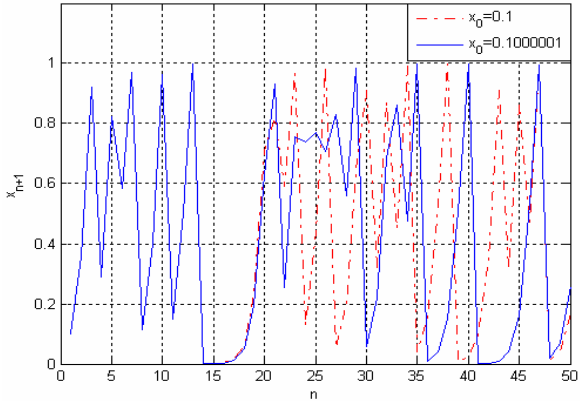

Fig. 5. Large variations in two trajectories for slightly different initial condition for μ = 4

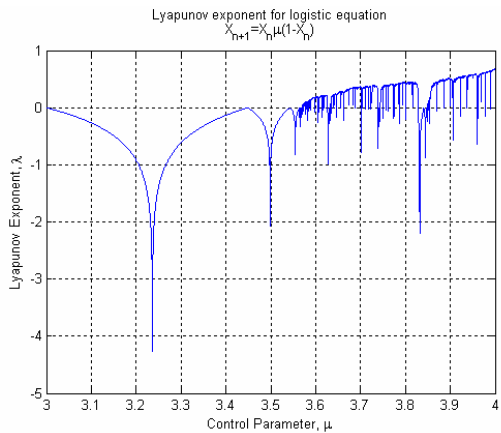

Fig. 6. Lyapunov Exponent

The is λ Lyapunov exponent which is given by

$$\lambda = \lim_{n \to \infty} \frac{1}{n} \sum_{i=0}^{n} \ln \left| f'(x_i) \right| \qquad (4)$$

Lyapunov exponent measure the rate at which nearby orbits converge or diverge. There are as many Lyapunov exponents as there are dimensions in the state space of the system, but the largest is usually the most important. The − ve value of λ shows that the orbit converge in time whereas + ve value of λ represent that the distance between nearby orbits grows in time and the system exhibits sensitive dependence on the initial condition. To show this we have experimentally found the following plot for Lyapunov exponent.

For this we have iterated (2) for μ = 3 to 4 with fine steps of 0.0001 with random initial conditions. The large sequence lengths of about 103000 out of which first 300 iterations are discarded in order to iterate the map long enough to allow transients to decay. The last current 1000 iterate are stored. Then Lyapunov exponent is calculated using (4). The procedure is repeated for various values of μ and plotted as shown in fig. 6. It is clearly evident from the plot that chaos occurs from μ = 3.57... to μ = 4 as Lyapunov exponent in this range are mostly positive.

Although most values beyond 3.57 exhibit chaotic behaviour, but there are still certain isolated values of μ that shows non-chaotic behavior; these are sometimes called *islands of stability*. For instance, around 3.82 (Refer fig. 7) there is a range of parameters μ which show oscillation between three values, and for slightly higher values of μ oscillation between 6 values, then 12 etc. There are other ranges where all oscillation periods do occur. At μ = 4 we found the value of Lyapunov exponent as 0.6932 which comes close to theoretical value i.e ln(2) ≈ 0.6931 with single precision.

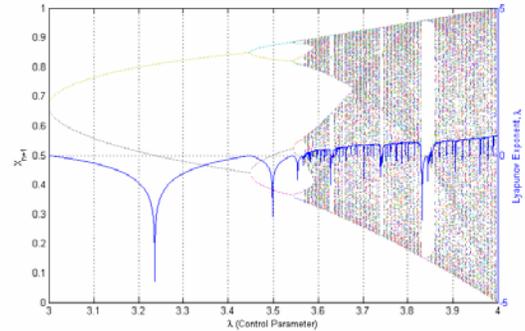

Fig. 7. Bifurcation Diagram and Lyapunov Exponent

The histogram of 10000 randomly chosen orbit of logistic map at μ = 4 is shown in fig. 8. We observe that $x_n$ values map to $x_{n+1}$ close to 1.0, which in turn map $x_{n+2}$ close to 0.0. Thus the probability density peaks at 1 and 0 with the distribution $p$ = $\dfrac{1}{\pi \sqrt{x(x-1)}}$

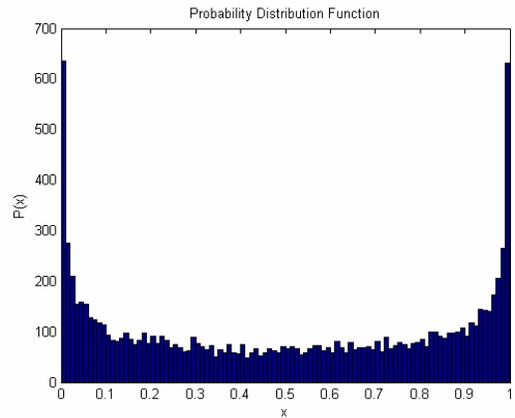

Fig. 8. PDF for randomly chosen orbit of 10000 at μ = 4

## III. RESULTS AND DISCUSSIONS

In order to test the feasibility and usefulness of the logistic map sequences for spread spectrum applications the following simulations were carried out.

As we know the chaotic sequence based on logistic map gives the real values which have to be binarised. This can be done by either of the following two ways:

$$f(x) = \begin{cases} 1 & x \geq 0.5 \\ 0 & x < 0.5 \end{cases} \tag{5}$$

$$f(x) = \begin{cases} 1 & x \geq \tau \\ 0 & x < \tau \end{cases} \qquad \text{where mean}[f(x)] = \tau \tag{6}$$

We have iterated both the sequences for various randomly chosen initial (seed) values. The result shows that the eq. (6) gives more balance of 1's and 0's as compare to eq. (5). Hence for all other simulations we have followed eq. (6). The autocorrelation for sample chaotic sequence for randomly selected initial condition (Fig. 9) shows very good correlation properties. Similarly Fig 10 shows the autocorrelation for chaotic sequences for varying period lengths and Fig 11 shows the cross-correlation of consecutive chaotic sequence for varying length of period.

## IV. Conclusion

We have shown the use of chaotic sequence in direct-sequence spread spectrum system as an alternative to PN sequence. The advantage of it being the availability of large number of variety of sequences of given length as compared to m-sequence and Gold sequence. The future work will be to analyse the effectiveness of chaotic sequence in watermarking applications which hides the secret information in any multimedia signal to protect copyright or prove authentication and so on.

## Appendix

### A. Proof of Lyaapunov exponent:

Start with initial condition $x_0$ and $x_0 + \delta_0$ ($\delta_0$ is very small)

Iterating the logistic map n steps:

$$x_n = f^n(x_0), \qquad x_n + \delta_n = f^n(x_0 + \delta_0)$$

The sequence $|\delta_n|$ is therefore,

$$|\delta_n| = |f^n(x_0 + \delta_0) - x_n|$$
$$= |f^n(x_0 + \delta_0) - f^n(x_0)|$$

and it gives the distance between two orbits for each n.
Now for small $\delta_n$,

$$|\delta_n| \approx |\delta_0| e^{\lambda n}$$

where $\lambda$ is called Lyapunov exponenet.

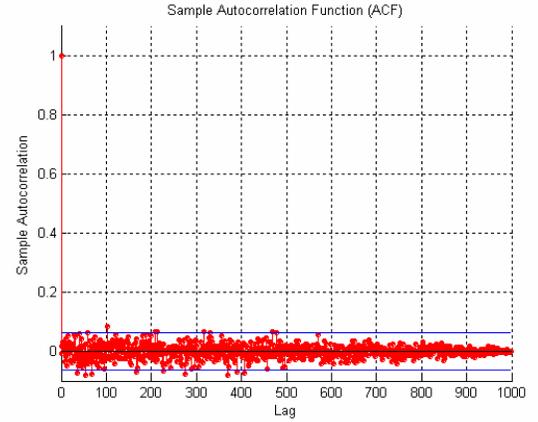

Fig. 9. Sample Autocorrelation for chaotic sequence

$$\Rightarrow \lambda \cong \frac{1}{n} \ln \left| \frac{\delta_n}{\delta_0} \right|$$

$$\cong \frac{1}{n} \ln \left| \frac{f^n(x_0 + \delta_0) - f^n(x_0)}{\delta_0} \right|$$

$$\cong \frac{1}{n} \ln \left| [f^n(x_0)]' \right|$$

However, the chain rule implies

$$[f^n(x_0)]' = [f(f^{n-1}(x_0))]'$$
$$= [f^{n-1}(x_0)]' f'(f^{n-1}(x_0))$$
$$= [f^{n-1}(x_0)]' f'(x_{n-1})$$

Repeating n-2 times

$$= [f'(x_0)(f'(x_1) f'(x_2) \dots f'(x_{n-1})]$$

So,

$$\lambda \cong \frac{1}{n} \ln \left| \prod_{i=0}^{n-1} f'(x_i) \right|$$

$$\cong \frac{1}{n} \sum_{i=0}^{n-1} \ln \left| f'(x_i) \right|$$

and now Lyapunov exponent can be defined more rigorously

$$\lambda = \lim_{n \to \infty} \frac{1}{n} \sum_{i=0}^{n} \ln \left| f'(x_i) \right|$$

Hence proved.

## References

[1] May R, "Simple mathematical models with very complicated dynamics," *Nature* (London), pp 261- 459, 1977.

[2] G. Heidari-Bateni, C.D. McGillem, "A Chaotic Direct-Sequence Spread-Spectrum Communication System," *IEEE Trans. on Comm.* Vol 42, 1524-1527, 1994.

[3] L. Kocarev, "Chaos based Cryptography: A brief Overview," Circuits and Systems Magazine IEEE, 2001.

[4] F. Dachselt and W. Schwarz, "Chaos and cryptography," *IEEE Trans. Circuits and Syst.* I, vol. 48, pp. 1498–1509, Dec. 2001.

[5] T. Stojanovski and L. Kocarev, "Chaos-Based random number generators—Part I: Analysis," *IEEE Trans. Circuits and Syst.* I, vol. 48, pp. 281 –288, Mar. 2001.

[6] Fridrich J, "Symmetric ciphers based on two-dimensional chaotic maps," J. Bifurcat Chaos 8 1259-84, 1998.

[7] I. J. Cox, M. L. Miller, and J. A. Bloom, "*Digital Watermarking,*" Morgan Kaufmann, London,

[8] I. J. Cox, J. Killian, T. Leighton, and T. Shamoon, "Secure spread spectrum watermarking for

[9] A. Nikolaidis and I. Pitas, "Comparison of different chaotic maps with application to image wat

[10] J. Fridrich, "Combining low-frequency and spread spectrum watermarking," *Proceedings SPIE*

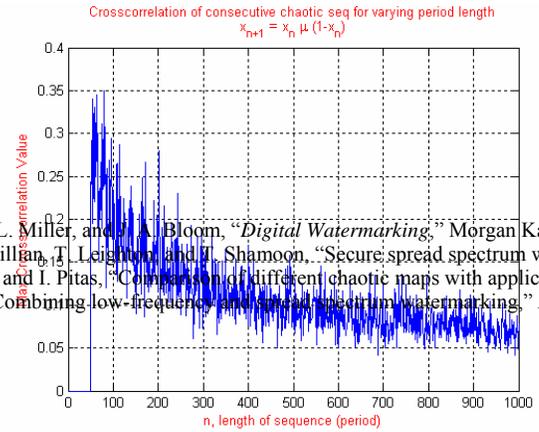

Fig. 11. Cross-correlation of consecutive chaotic sequence for varying period length

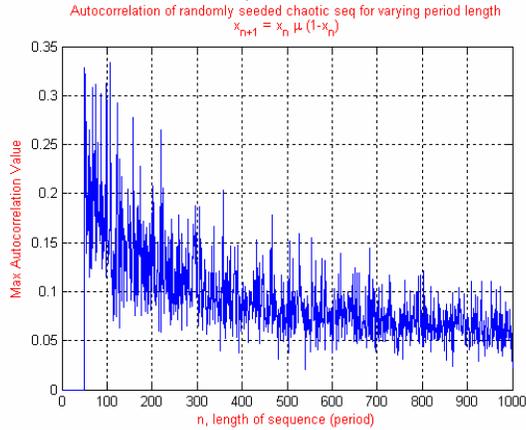

Fig. 10. Autocorrelation of randomly seeded chaotic sequence for varying period length